RESEARCH ARTICLE

# Gender Differences in Appropriate Shocks and Mortality among Patients with Primary Prophylactic Implantable Cardioverter-Defibrillators: Systematic Review and Meta-Analysis

David Conen[1]*, Barbora Arendacká[2], Christian Röver[2], Leonard Bergau[3], Pascal Munoz[3], Sofieke Wijers[4], Christian Sticherling[5], Markus Zabel[3], Tim Friede[2,6]

1 Division of Internal Medicine, Department of Medicine, University Hospital Basel, Basel, Switzerland,
2 Department of Medical Statistics, University Medical Center Göttingen, Göttingen, Germany,
3 Department of Cardiology and Pulmonology and Clinical Electrophysiology Division, University Medical Center Göttingen, Göttingen, Germany, 4 Department of Physiology and Cardiology, Universitair Medisch Centrum Utrecht, Utrecht, The Netherlands, 5 Division of Cardiology, Department of Medicine, University Hospital Basel, Basel, Switzerland, 6 German Centre for Cardiovascular Research (DZHK), partner site Göttingen, Göttingen, Germany

* david.conen@usb.ch

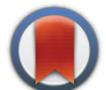

Data Availability Statement: This is a meta-Analysis of previously published data. All relevant data are within the paper and its Supporting Information files.

Funding: The research leading to the results has received funding from the European Community's Seventh Framework Program FP7/2007-2013 under grant agreement№ 602299, EU-CERTICD. David Conen was supported by grants of the Swiss National Science Foundation (PP00P3_159322). The funders had no role in study design, data collection and

## Abstract

### Background

Some but not all prior studies have shown that women receiving a primary prophylactic implantable cardioverter defibrillator (ICD) have a lower risk of death and appropriate shocks than men.

### Purpose

To evaluate the effect of gender on the risk of appropriate shock, all-cause mortality and inappropriate shock in contemporary studies of patients receiving a primary prophylactic ICD.

### Data Source

PubMed, LIVIVO, Cochrane CENTRAL between 2010 and 2016.

### Study Selection

Studies providing at least 1 gender-specific risk estimate for the outcomes of interest.

### Data Extraction

Abstracts were screened independently for potentially eligible studies for inclusion. Thereby each abstract was reviewed by at least two authors.

## Data Synthesis

Out of 680 abstracts retained by our search strategy, 20 studies including 46'657 patients had gender-specific information on at least one of the relevant endpoints. Mean age across the individual studies varied between 58 and 69 years. The proportion of women enrolled ranged from 10% to 30%. Across 6 available studies, women had a significantly lower risk of first appropriate shock compared with men (pooled multivariable adjusted hazard ratio 0.62 (95% CI [0.44; 0.88]). Across 14 studies reporting multivariable adjusted gender-specific hazard ratio estimates for all-cause mortality, women had a lower risk of death than men (pooled hazard ratio 0.75 (95% CI [0.66; 0.86]). There was no statistically significant difference for the incidence of first inappropriate shocks (3 studies, pooled hazard ratio 0.99 (95% CI [0.56; 1.73]).

## Limitations

Individual patient data were not available for most studies.

## Conclusion

In this large contemporary meta-analysis, women had a significantly lower risk of appropriate shocks and death than men, but a similar risk of inappropriate shocks. These data may help to select patients who benefit from primary prophylactic ICD implantation.

## Introduction

Several landmark studies have shown that the primary prophylactic use of implantable cardioverter defibrillators (ICD) among patients with reduced left ventricular ejection fraction is associated with a significant reduction in all-cause mortality [1–3]. These findings were rapidly adopted by guidelines and have since then become standard of care in this patient population.

However, ICDs are costly, can lead to inappropriate ICD therapy and put a heavy burden on the healthcare system [4, 5], providing an impetus for better risk stratification for primary prophylactic ICD implantation. Medical treatment for patients with heart failure and reduced ejection fraction has considerably improved since the publication of the randomized trials for primary prevention ICD implantation [6]. In addition, the proportion of lower risk patients with non-ischemic cardiomyopathy among ICD recipients continues to increase. Thus, the overall benefit from these devices may be lower than initially estimated. This may be particularly true for subgroups at lower risk of sudden cardiac death.

Some but not all recent studies have suggested that women may have a lower risk of sudden cardiac death than men [7–9], suggesting that gender may be an easily determinable factor to be considered for risk stratification. Two meta-analyses of randomized trials concluded that women had either no benefit or a smaller benefit than men [10, 11]. Some studies also suggested a higher risk of complications in women, further underscoring the potential importance of considering gender when balancing risks and benefits of primary prophylactic ICD implantation.

We therefore performed a systematic review and meta-analysis of contemporary studies to assess gender related differences in outcomes among patients undergoing primary prophylactic ICD implantation.

## Methods

### Search strategy

The aim of this meta-analysis was to synthesize published results from contemporary studies regarding the effect of gender on the risk of appropriate shock, all-cause mortality and inappropriate shock in patients with ICD implanted for primary prevention. Accordingly, we searched PubMed, LIVIVO and Cochrane CENTRAL (date of last search: May 11, 2016) for relevant studies published from 2010 onward using the following search terms: ("primary prophylaxis" OR "primary prophylactic" OR "primary prevention") AND ("ICD" OR "defibrillator") AND ("mortality" OR "shock" OR "death" OR "ICD therapy" OR "ICD treatment"). No language restrictions were applied to the search. We did not consider meeting abstracts or other gray literature. The year 2010 was chosen as a starting date in order to limit the search to studies that had enrolled predominantly primary prophylactic ICD patients after the publication of major landmark trials and corresponding guidelines in the field [1–3]. However, the patient cohorts might include patients with ICD implantation before 2010. The yield of our search strategy was checked against a pre-defined list of 19 publications that are related to the topic and that we had compiled prior to the search. Reference lists of all publications fulfilling the inclusion criteria were also screened to identify additional publications.

### Study selection

The abstracts identified by the literature search as described in the previous paragraph were reviewed independently and each abstract was seen by at least two authors (LB, PM, MZ, BA). If an abstract was judged as potentially relevant by at least one of the reviewers, the full-text of the publication was screened 1) for appearance of at least one of the three endpoints of interest appropriate shock, all-cause death or inappropriate shock, and 2) for reported gender-specific effects on at least one of the end-points. Further, we required the study population to be limited to patients with an implanted ICD, with or without cardiac resynchronization therapy (CRT), and who were enrolled at the time of device implantation, in order to minimize the risk of survival bias. To focus our results on patients with an ICD implanted for primary prevention, we considered only papers in which at least 60% of the study population received a primary prophylactic ICD, or in which results for the primary prophylactic subgroup were reported separately. In 2 cases we had access to the individual patient data and re-analyzed the data within the primary prevention subgroup [12, 13]. Papers considering very specific patient populations, e.g. all patients with CRT-D, all patients older than 80 years, patients after CABG surgery only or all patients on dialysis, were excluded.

### Assessment of study quality

Since this systematic review is based on observational studies only (and does not include randomized controlled trials), we assessed the risk of bias regarding three domains (selection of participants, measurement of variables and outcomes, control of confounding) specific to observational studies as previously recommended [14].The study quality was assessed by two reviewers (BA, CR).

### Data extraction

Hazard ratios quantifying the effect of gender were extracted from univariable or multivariable Cox proportional hazards models, and constituted the effect measure of interest. Specifically, we extracted point estimates, their standard errors (when available) and associated confidence intervals (CI). The text was screened for consistent reporting of the results, in order to guard against possible typographical errors. To the same end, the reported intervals were checked for symmetry around the point estimate on the log-scale. If inconsistencies were observed or questions regarding variable definitions arose, we contacted the corresponding authors of the respective publications and asked for clarification. In cases where individual patient data were available and the original paper did not report hazard ratios for the primary prophylactic patients, we obtained the needed hazard ratios by repeating the analysis from the respective publication using the appropriate data subset.

In addition to hazard ratios, we extracted information on the total number of patients included, duration of follow up, year of ICD implantation and on a pre-defined list of baseline characteristics, including gender, age, ischemic cardiomyopathy, NYHA functional class, left ventricular ejection fraction (LVEF), creatinine concentration and/or glomerular filtration rate, QRS duration, diabetes mellitus, type of device and primary versus secondary prevention. Data extracted by one author were independently verified by another author.

### Statistical analysis

The extracted hazard ratios were log-transformed and their standard errors, if not available directly, were calculated from the reported 95% CIs [15]. Random effects models using Mandel-Paule estimators of the between-study variance were applied to pool the log-transformed hazard ratios. The 95% CI for the pooled effect estimate was calculated using the Knapp-Hartung approach with the suggested ad hoc adjustment [16]. Between-study heterogeneity was assessed by the Cochran Q chi-square test and by the $I^2$ measure (as implemented in the R package *metafor* [17]). The stability of the results was examined by leaving out one study at a time and re-pooling the remaining hazard ratios. All analyses were done using the R software (R Foundation for Statistical Computing, Vienna, Austria). A p-value <0.05 was pre-specified to indicate statistical significance.

## Results

### Search results

After removing duplicates, the search yielded a list of 680 potentially relevant records. Out of these, 264 abstracts were identified for full-text screening (Fig 1). The absence of gender-specific effect estimates on at least one of the endpoints and exclusion of studies on very specific patient populations narrowed the number of potentially eligible publications to 43. Additional reasons for exclusions are listed in Fig 1. Results from 1 manuscript [18] were excluded because of an indirect classification of the prevention type, and because the underlying sample population overlapped to a large extent with a more precisely defined primary prevention group from another study (9).

Thus, 20 papers qualified for our quantitative synthesis [7–9, 12, 13, 19–33], two of them [26, 28] from the same study population. For the meta-analysis on mortality, the results reported by Yung et al were considered since their report focusses on the primary prevention subgroup [28]. For the analyses reporting on appropriate and inappropriate shocks we included the effect estimates from MacFadden et al. [26] in our analyses, since standard Cox proportional hazards analyses were not provided in the former publication [28].

An overview of the 20 included studies is shown in Table 1 and more detailed characteristics are presented in Tables 2 and 3. Overall, the included studies provided results on up to 46'657 patients with gender-specific information on at least one of the relevant endpoints. Mean follow-up across the individual studies ranged from 0.78 to 5.4 years. Mean age varied between 58 and 69 years. The proportion of women enrolled ranged from 10% to 30%. The prevalence of

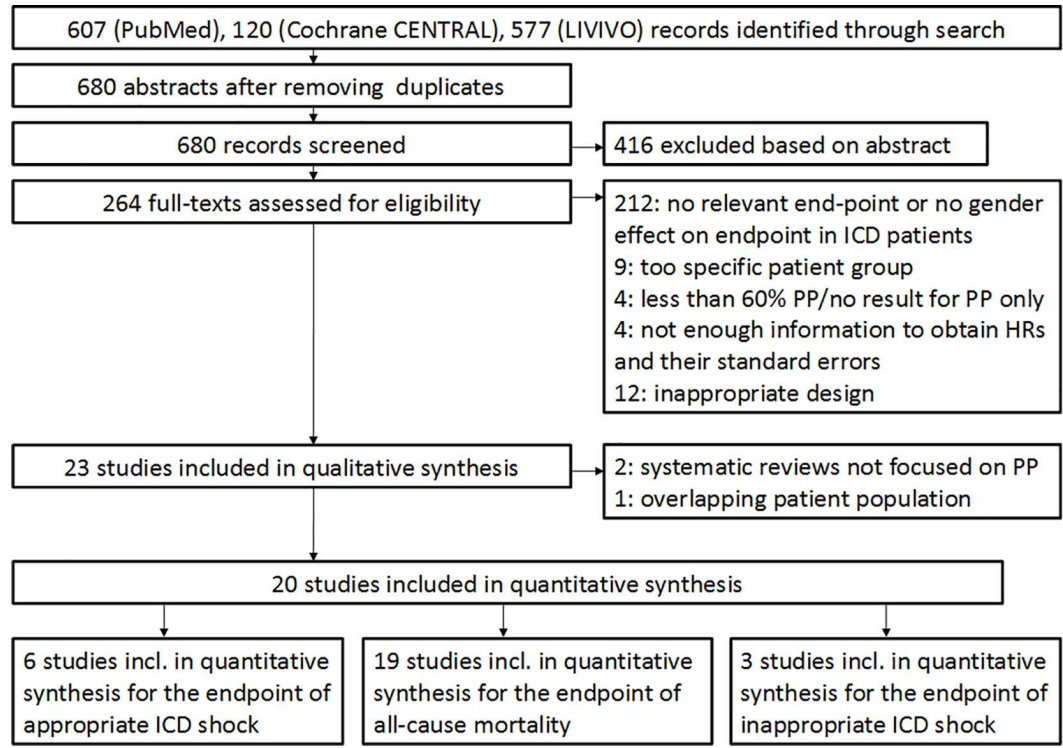

**Fig 1. Flow diagram of the systematic review.** PP = primary prevention, HR = hazard ratio.

doi:10.1371/journal.pone.0162756.g001

ischemic cardiomyopathy was between 56% and 100%. All but four studies provided results on primary prevention patients only.

## Assessment for bias

All studies included in the meta-analyses recruited consecutive patients undergoing ICD implantation, in one case [20] the study population was limited to recipients of dual chamber ICDs without prior atrial fibrillation, in one case [25] the population of patients with available NT-proBNP or BNP measurements was complemented by a parallel cohort without these measurements. All studies provided description of the inclusion/exclusion criteria and matched our objectives. All studies provided details on the determination of the end-points, however, only 9 [7, 12, 13, 20–22, 27–29] discussed to some extent measurement of baseline variables. In all but 5 cases related to all-cause mortality the statistical analyses accounted for confounders, see also Table 4. We analyzed these 5 studies separately (Fig 2). We did not exclude any of the identified studies from further analysis after assessment of bias risk.

## Appropriate shocks

Effect of gender on the incidence of first appropriate shock was available from 6 studies (Table 1). Covariates of the multivariable models for the incidence of first appropriate shock are shown in Table 4. After combining the 6 individual hazard ratios, women had a significantly lower risk of receiving an appropriate shock than men, with a pooled risk estimate of 0.62 (95% CI [0.44; 0.88], p = 0.0175) (Fig 3). The between study standard deviation was 0.20 and $I^2$ was 36% with a p-value for heterogeneity of 0.2381. Removing the study by Weeke et al. eliminated the between study heterogeneity, without significantly influencing the relative risk

Table 1. Overview and available endpoints of eligible papers.

| Study | N patients | Implantation period | PP [%] | Appropriate shock | All-cause mortality | Inappropriate shock |
|---|---|---|---|---|---|---|
| van der Heijden et al | 1946 | 1996–2012 | 100 | yes | yes | yes |
| Seegers et al* | 632 | 2000–2010 | 100 | yes | yes | — |
| Weeke et al | 1609 | 2007–2011 | 100 | yes | yes | yes |
| Wijers et al* | 553 | 2006–2011 | 100 | yes | yes | — |
| Yung et al* | 3939 | 2007–2010 | 100 | —[&] | yes | — |
| Gatzoulis et al[†] | 422/495 | 1992–2010 | 78.4/76.3 | yes | yes | — |
| MacFadden et al | 5213 | 2007–2010 | ≈70 | yes | yes[#] | yes |
| Bilchick et al | 17991 | 2005–2006 | 100 | — | yes | — |
| Gigli et al | 193 | 2003–2010 | 100 | — | yes | — |
| Hage et al* | 409 | 2002–2007 | 100 | — | yes | — |
| Masoudi et al | 2954 | 2006–2010 | 100 | — | yes | — |
| Providência et al | 5539 | 2002–2012 | 100 | — | yes | — |
| Rodríguez-Mañero et al | 1174 | 2008–2011 | 100 | — | yes | — |
| Smith et al | 427 | 2004–2009 | 100 | — | yes | — |
| Amit et al | 1518 | 2010–2013 | ≈70[§] | — | yes | — |
| Campbell et al | 197 | 2003–2009 | 100 | — | yes | — |
| Demirel et al | 94 | 2004–2010 | 100 | — | yes | — |
| Kraaier et al | 861 | 2002–2008 | 100 | — | yes | — |
| Levine et al | 783 | 2003–2012 | 100 | — | yes | — |
| Stabile et al | 130 | 2002–2003 | >70 | — | yes | — |

PP = Primary prevention

* Primary prevention subgroup

[†] Subgroup of patients with LVEF< = 35%/all patients

[§] Indirect estimate based on the 74% of PP in the baseline cohort (3543 patients) and the claim of no significant baseline differences between patients with and without follow-up

[#] Not included in the meta-analysis, given a large overlap with Yung et al

[&] Appropriate shock considered, but hazard ratio from a Cox proportional hazards model not provided.

doi:10.1371/journal.pone.0162756.t001

of appropriate shocks (hazard ratio 0.69, 95% CI [0.52; 0.92], $p$ = 0.0222). Similarly, excluding any one study from the meta-analysis had no appreciable effect on the overall results, as shown in S1 Table.

### All-cause mortality

A gender specific risk estimate for all-cause mortality was available in 19 studies. Out of the 19 available hazard ratios, 14 are based on multivariable (covariates listed in Table 4) and 5 on univariable models. Women had a lower risk of death than men (hazard ratio 0.78, 95% CI [0.68; 0.89], $p$ = 0.001), as shown in Fig 2. The between study standard deviation was 0.1544 (heterogeneity test p = 0.0436) and $I^2$ was 47%, suggesting moderate between-study heterogeneity. Excluding two small studies [20, 27] reporting extremely wide CIs did not materially influence the pooled risk estimate (0.79, 95% CI [0.71; 0.87], $p$ = 0.0001), but reduced between study heterogeneity (between study standard deviation 0.0770, $I^2$ = 20%, heterogeneity test p = 0.1171). Again, results were stable and not sensitive to the exclusion of any one study from the meta-analysis, as shown in S1 Table for studies reporting multivariable adjusted hazard ratios.

Table 2. Baseline characteristics of the selected studies.

| Baseline characteristic | van der Heijden et al | Seegers et al (PP only) | Weeke et al | Wijers et al (PP only) | Yung et al (PP only) | Gatzoulis et al (LVEF≤35%) | MacFadden et al | Bilchick et al. | Gigli et al. | Hage et al. (PP only) |
|---|---|---|---|---|---|---|---|---|---|---|
| n | 1946 | 632 | 1609 | 553 | 3939 | 422 | 5213 | 17991 | 193 | 409 |
| Follow-up [y] | 3.3 (1.4–5.4) | 4.2±2.1 | 1.9±1.3 | 2.4±1.57 | 1.8 (1.0–2.8) | 3.3±2.92* | 0.78* (mean) | up to 5 years | 4.2* (mean) | 4.2±2** |
| Male | 78.5% | 82% | 84.2%* | 72% | 80.2%* | 90% | 78.8%* | 77.5%* | 84% | 70.2%* |
| Age [y] | 65±15 | 65.5±12 | M:67.5 (60.7–73.2) F:69.0 (61.0–74.6) | 63.3±11.4 | 65.2 ±10.75* | 65.6±10.2 | n.g.§ | <74: 61.2%* ≥75: 38.8%* | 64.4 ±10 | 58 ±14.2* |
| ICM | 66% | 67.2% | 100% | 61.8% | 70.4%* | 69.4% | n.g. | n.g. | 62% | n.g. |
| NYHA | | | | | | 2.6±0.6 | | | n.g. | |
|    I | 18.3% | 14.7% | 6.2%* | 6.9% | 23.1%* | | n.g.§ | 8.8% | | 14.7%* |
|    II | 34.6% | 24.7% | 50.5%* | 26.6% | 40.7%* | | n.g.§ | 51.0% | | 37.2%* |
|    III-IV | 44.8%* | 59% | 38.3%* | 42% | 36.2%* | | n.g.§ | 40.2% (III) | | 43.7%* |
|    unknown | 2.3% | 1.6% | 5%* | 24.6% | 0% | | | 0% | | 4.4%* |
| LVEF [%] | 29±11.7 | 27±8.6‡ | 25(20–30)* | 23.5±6.31 | ≤20: 21.1%* 20–30: 55.5%* ≥31: 18.4%* NA: 5%* | 26.9±15.5 | n.g.§ NA: 1039 of 5213 | ≤20: 31.6% | 26±6 | 26±13* |
| Creatinine (C)/ eGFR[mL/min/ 1.73m²] | C: 78.7 ±36.4 ml/ min | eGFR: 65.9 ±23.3‡ | n.g. | eGFR: 63.7 ±26.7 C:114.2 ±61.7 µmol/L | C:111.2 ±62.9* µmol/L | n.g. | n.g.§ | n.g. | n.g. | eGFR: 70±23 |
| QRS [ms] | 132±35.9 | 129±37‡ | SC: 100(90–115) DC: 100(90–120) CRT-D: 150 (130–168) | 135.6±32.8 | 134.0 ±35.53* | n.g. | n.g.§ | <120: 59.0% 120–149: 22.2% ≥150: 18.8% | n.g. | 133±34 |
| Diabetes | 23% | 27.2% | n.g. | n.g. | 37%* | n.g. | n.g.§ | 33.6% | n.g. | 33.5% |
| Primary prev. | 100% | 100% | 100% | 100% | 100% | 78.4%* | 70%** | 100% | 100% | 100% |
| Type of device | | | | | | n.g. | n.g.§ | n.g. | | n.g.§ |
| SC | 4% | 29.1% | 46.2% | 45.9% | 43.5%* | | | | 72.5%* | |
| DC | 38% | 25.6% | 16.8% | 7.1% | 23.5%* | | | | 27.5%* | |
| CRT-D | 58% | 45.3% | 37.0% | 47.0% | 33%* | | | | 0% | |

The values are given as percentages, mean±SD or median (IQR). PP = primary prevention, M = male, F = female, SD = standard deviation,
IQR = interquartile range, n.g. = not given, SC = Single chamber, DC = Dual chamber, CRT-D: Cardiac resynchronization therapy-defibrillator.
NA = unknown, y = year.
* calculated
** calculated from 5450 pts undergoing ICD implantation (237 later excluded due to the lack of follow-up)
§ numbers given only for a larger group of pts referred to ICD implantation
‡ calculated from non-missing values

doi:10.1371/journal.pone.0162756.t002

## Inappropriate shock

Effect of gender on the risk of first inappropriate shock was available in 3 studies. Gender had no apparent effect on the occurrence of inappropriate shocks, with a pooled hazard ratio of 0.99 (95% CI [0.56–1.73], p = 0.9276) and no evidence for heterogeneity ($I^2$ = 0, heterogeneity test p = 0.47), as shown in Fig 4. Excluding any one study from the meta-analysis had no appreciable effect on the overall results (see S1 Table).

Table 3. Baseline characteristics of the selected studies.

| Baseline characteristic | Masoudi et al | Providência et al | Rodríguez-Mañero et al | Smith et al | Amit et al | Campbell et al | Demirel et al | Kraaier et al | Levine et al | Stabile et al |
|---|---|---|---|---|---|---|---|---|---|---|
| n | 2954 | 5539 | 1174 | 427 | 1518 | 197 | 94 | 861 | 783 | 130 |
| Follow-up [y] | 2.4(1.3–3.8) | ≈2.7 (med) | 3.2±1.8 | 2.6(1.25–3.75)* | 0.88 (med) | 2.8 (med) | 5.4(4.5–6.6) | up to 1 y | 3.6±3.1 | 5.25±1 |
| Male | 74% | 84.9%* | 81.4%* | 79% | 83%* | 85.8% | 86.2% | 78.7% | 78.4% | 77%┤ |
| Age [y] | 69(60–75) | 62.5±11.2 | 62.7±11.1 | 58±14 | n.g.& | 66.7±9.6 | 65±10.7 | 62.7±10.2 | n.g.# | 66±9 ┤ |
| ICM | 62.2% | 59.6%* | 56%* | 68% | n.g.& | 100% | 100% | 67.1%*‡ | n.g.# | 56% ┤ |
| NYHA | | | | | n.g.& | | | | n.g.# | |
|     I | | 8.7%* | n.g. | | | n.g. | 36.2%* | | | 3% ┤ |
|     II | 61.2%(I+II) | 33.6%* | n.g. | 81.3% (I-II) | | n.g. | 42.6%* | 67.6%* (I-II) | | 23% ┤ |
|     III- IV | 38.6% | 40.3%* | 38.8%* | 18% | | 64.5%* | 19.1% * (III) | 32.3% * | | 74% ┤ |
|     unknown | 0.2% | 17.4%* | n.g. | 0.7% | | 11.7%* | 2.1%* | 0.1%* | | 0% ┤ |
| LVEF [%] | ≤30: 85.4% 31–35:14.6% | 26.7±7.2 | 26.2±7.6 | 27±9 | n.g.& | 25.8(20.0–30.0)‡ | 31.9±9.3 | 24.3±8.7‡ | n.g.# | 30±9 ┤ |
| Creatinine (C)/ eGFR[mL/min/ 1.73m$^2$] | C: 1.4±0.9 mg/dL eGFR: 61.5±22.5 | eGFR: <60: 23%* | C: 1.18±0.6 mg/dL | C: 97 ±41 µmol/L eGFR: 78 ±26 | n.g.& | eGFR: ≥90: 8.1%, 60–89: 35.5%, 30–59: 49.7%,<30: 6.6% | n.g. | eGFR: 74.4 ±63.3‡ | n.g.# | n.g. |
| QRS [ms] | >120: 49.1% | <120: 21.4%* | 127.7±32.2 | 116±26 | n.g.& | n.g. | n.g. | 127±33‡ | n.g.# | 138±34 ┤ |
| Diabetes | 42.2% | n.g. | 33.0% | 21% | n.g.& | 32% ‡ | 22.3%* ‡ | 18.2%*‡ | n.g.# | 20% ┤ |
| Primary prev. | 100% | 100% | 100% | 100% | ≈70%□ | 100% | 100% | 100% | 100% | 74% ┤ |
| Type of device | | | | | n.g.& | | n.g. | n.g. | n.g.# | |
| SC | 32.3% | 22.7% | 38.9% | 74% | | 0% | | | | n.g. |
| DC | 36.0% | 23.1% | 18.9% | 26% | | 100% | | | | n.g. |
| CRT-D | 31.7% | 53.3% | 42.2%* | 0% | | 0% | | | | 65% ┤ |

The values are given as percentages, mean±SD or median (IQR). Med = median. M = male, F = female, SD = standard deviation, IQR = interquartile range, n.g. = not given, SC = Single chamber, DC = Dual chamber, CRT-D: Cardiac resynchronization therapy-defibrillator.

* calculated
** in a larger cohort including secondary prevention patients
§ 42.5% pts had ICD with biventricular pacing device
& not given for the subgroup of patients with follow-up
□ see note under Table 1
‡ missing values excluded
# not given for the larger cohort underlying all-cause mortality results
┤ in a larger cohort of 139 patients (for 9 missing information on survival status)

doi:10.1371/journal.pone.0162756.t003

## Discussion

In this meta-analysis of contemporary studies among patients receiving a primary prophylactic ICD because of a depressed left ventricular ejection fraction, we found that women have a lower incidence of a first appropriate shock and death than men, but a similar risk of receiving inappropriate shocks. These results provide important insights on the risk-benefit ratio in specific subpopulations eligible for primary prophylactic ICD implantation.

Some of the included studies included a significant proportion of patients with ICD implantation for secondary prophylaxis. To investigate whether this might potentially bias any

**Table 4. List of covariates for which the reported hazard ratios were adjusted.**

| | |
|---|---|
| **van der Heijden et al** | - **AS, M**: age, aetiology of heart failure, device type, LVEF, NYHA class, history of atrial fibrillation/flutter, creatinine clearance, usage of β-blockers<br>- **IAS**: atrial fibrillation/flutter |
| **Seegers et al** | - **AS**: age, Amiodarone;<br>- **M**: age, eGFR, diuretics, peripheral arterial disease |
| **Weeke et al** | - **AS, IAS**: age, QRS duration, LVEF, type of device (time dependent), history of percutaneous intervention, history of CABG, implantation year, atrioventricular conduction disease, device upgrade<br>- **M**: as for AS, plus therapy during follow up (appropriate/inappropriate shock, appropriate/inappropriate ATP) |
| **Wijers et al** | - **AS:** LVEF, ischemic cardiomyopathy<br>- **M**: LVEF, QRS duration, GFR |
| **Yung et al** | - **M**: age, NYHA class, syncope, peripheral vascular disease, GFR, left atrial size, prescribed ACE inhibitors or ARB, prescribed loop diuretics |
| **Gatzoulis et al** | - **AS**: age, ICM, prevention type<br>- **M**: age, ischemic cardiomyopathy, LVEF< = 35%, NYHA, type of prevention |
| **MacFadden et al** | - **AS, IAS**: age, QRS duration, creatinine, hemoglobin, systolic and diastolic blood pressure (each squared), NYHA, ventricular tachycardia or fibrillation or nonsustained ventricular tachycardia, myocardial infarction, previous percutaneous coronary intervention or CABG, heart failure, family history of SCD, atrial fibrillation, valvular heart disease, syncope, dyslipidemia, diabetes mellitus, hypertension, previous stroke or transient ischemic attack, peripheral vascular disease, COPD, LVEF, device type |
| **Bilchick et al** | - **M**: age, race, QRS duration, bundle branch block, atrial fibrillation, LVEF, NYHA, duration of heart failure, diabetes mellitus, COPD, chronic kidney disease, prior myocardial infarction, prior CABG, systolic blood pressure, diastolic blood pressure, heart rate, digoxin, beta-blockers, ACE inhibitors, diuretic agents, Amiodarone, Warfarin, breast cancer, colon cancer, prostate cancer, depression |
| **Gigli et al** | - **M**: age, LVEF, type of device, ischemic cardiomyopathy |
| **Hage et al** | - **M**: age, hypertension, atrial fibrillation, myocardial infarction, LVEF, left bundle branch block, biventricular pacing, Amiodarone, other antiarrhythmics, β-blockers, chronic kidney disease |
| **Masoudi et al** | - **M**: LVEF, ischemic cardiomyopathy, NYHA, blood urea nitrogen, atrial fibrillation, diabetes, hypertension, chronic lung disease, hemoglobin, QRS, device type, ACE/ARB therapy, β-blockers |
| **Providência et al.** | - **M**: NYHA, atrial fibrillation, ischemic cardiomyopathy, QRS, CRT-D, β-blockers, Amiodarone, spironolactone, calcium channel blockers, antiplatelet agents, vitamin K antagonists |
| **Rodríguez-Mañero et al.** | - **M**: LVEF, age, creatinine, COPD, digoxin therapy |
| **Smith et al** | - **M**: age, NYHA, diuretic use, ACE inhibitor use, renal failure |
| **Amit et al** | - **M**: age, device type, LVEF, prevention type, diabetes, β-blockers, renal function |

AS = appropriate shock, M = all-cause mortality, IAS = inappropriate shock.

doi:10.1371/journal.pone.0162756.t004

conclusions for primary prophylactic ICD implantation we performed sensitivity analyses including only studies completely focused on primary prevention. The combined effects were very similar to the findings reported here providing reassurance in the approach taken.

The proportion of participants with ischemic cardiomyopathy in the included cohorts was relatively high, a group of patients with a higher mortality risk than those with non-ischemic cardiomyopathy. In the MADIT-II study, where only patients after a myocardial infarction were enrolled, mortality rate at 2 years was approximately 15% in the intervention group [2]. SCD-Heft enrolled a mixed population of ischemic and non-ischemic cardiomyopathies and found a risk of death at 2 years in the intervention group of approximately 11–12% [1]. Estimated 2-year mortality rates from the cohorts included in this analysis are shown in the

PLOS ONE | Gender Differences in Patients with Primary Prophylactic Implantable Cardioverter-Defibrillators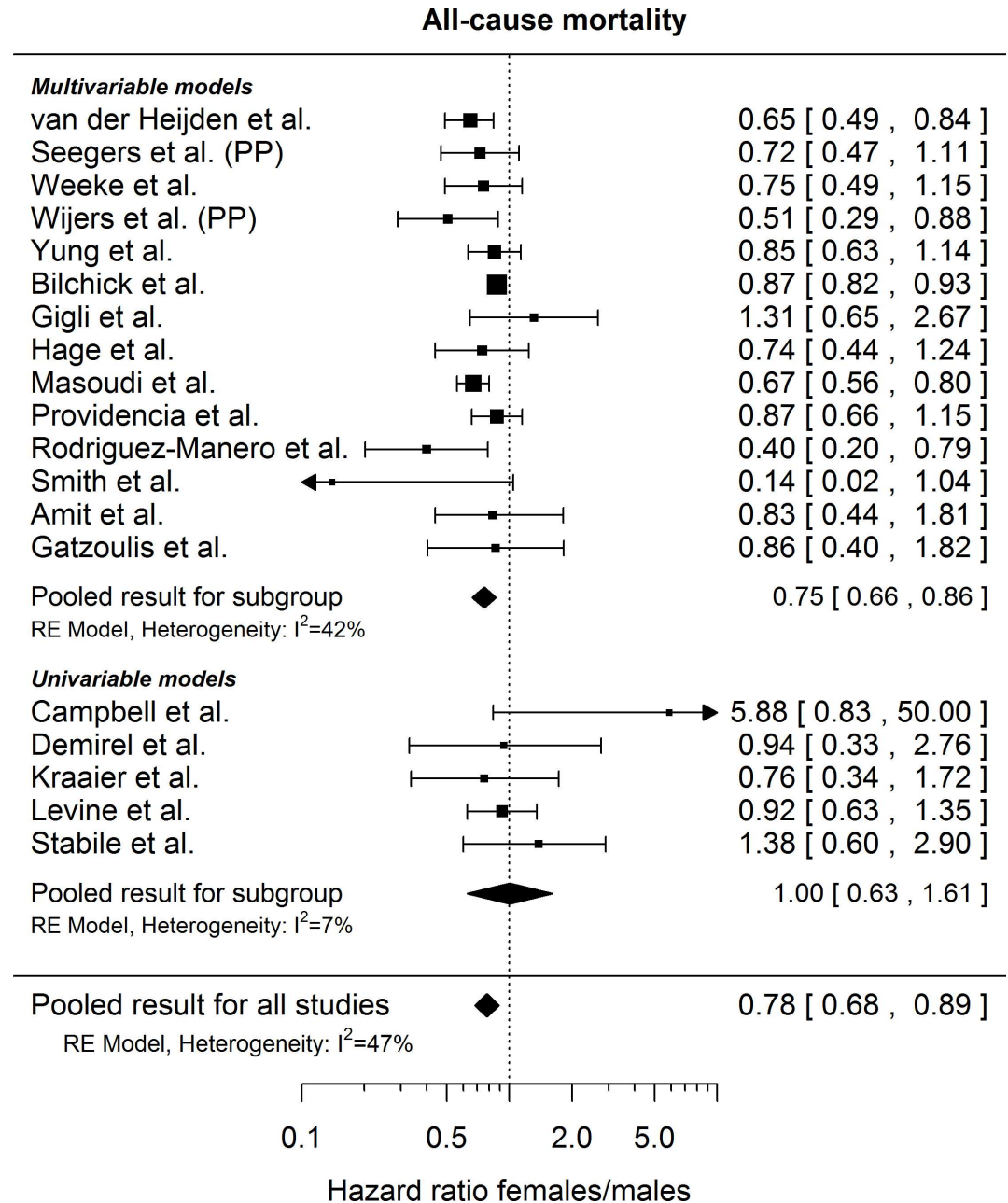

**Fig 2. Extracted hazard ratios for female gender regarding risk of appropriate shocks with 95% confidence intervals as reported in the respective publications.** 'PP' indicates that the results were re-analyzed for primary prevention patients only. The pooled estimate is reported with a Knapp-Hartung adjusted 95% confidence interval. The dotted vertical line denotes a hazard ratio of 1, which corresponds to no difference in the risk between males and females.

doi:10.1371/journal.pone.0162756.g002

S2 Table. These rates obtained from contemporary cohort studies are comparable to those in ScD-Heft or even slightly lower, potentially showing the improvements of medical treatment among patients with heart failure and reduced ejection fraction over the last 15 years [34, 35]. Assuming a similar relative benefit of primary prophylactic ICD implantation despite lower

PLOS ONE | DOI:10.1371/journal.pone.0162756  September 12, 2016  10 / 15

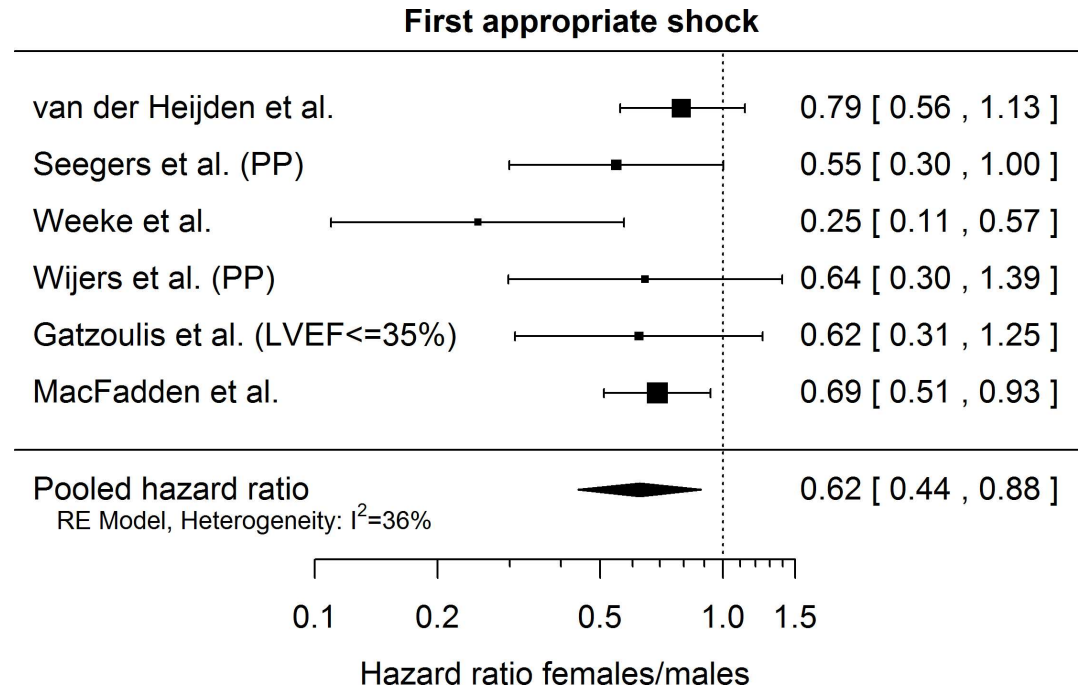

**Fig 3. Extracted hazard ratios for female gender regarding risk of all-cause mortality with 95% confidence intervals as reported in the respective publications.** 'PP' indicates that the results were re-analyzed for primary prevention patients only. The pooled estimate is reported with a Knapp-Hartung adjusted 95% confidence interval. The dotted vertical line denotes a hazard ratio of 1, which corresponds to no difference in the risk between males and females.

doi:10.1371/journal.pone.0162756.g003

mortality rates, the absolute benefit of the device decreases in the context of an improved survival in this patient population.

These are important considerations, as ICDs are costly and their implantation is associated with the occurrence of several important complications such as inappropriate shocks, infections and lead malfunction. In an earlier study, the rate of major complications in the first year after ICD implantation was 13.9 per 100 person-years among women and 7.4 per 100 person-years among men, with an adjusted odds ratio of 1.91 (1.48–2.47; p<0.001) [26]. In the current meta-analysis we showed that women had a 22% lower risk of death and a 38% lower risk of appropriate shocks compared with men but a similar risk of inappropriate shocks.

Thus, the available evidence suggests that the risk-benefit ratio might be less favorable in women and the number needed to save one life higher. While our findings do certainly not mean that primary prophylactic ICD therapy should be withheld in women, they do suggest that in the context of diminishing absolute benefits and growing costs in most health care systems that improved risk stratification tools for primary prophylactic ICD implantation in patients with heart failure are urgently needed and are likely to include gender as a prognostic factor. Prior meta-analyses of randomized trials have shown that women have a similar risk of death but a lower risk of appropriate shocks compared with men [11], suggesting that at least the lower risk of appropriate shocks found in our study is not entirely due to confounding. Taken together, improved risk stratification is needed to allocate primary prophylactic ICD treatment, and gender may be one of many possible risk factors, such as life expectancy, chronic kidney disease, atrial fibrillation or chronic obstructive pulmonary disease [19, 36].

Strengths of this comprehensive meta-analysis include the large sample size and the focus on contemporary patients treated after the publication of the landmark studies in the field.

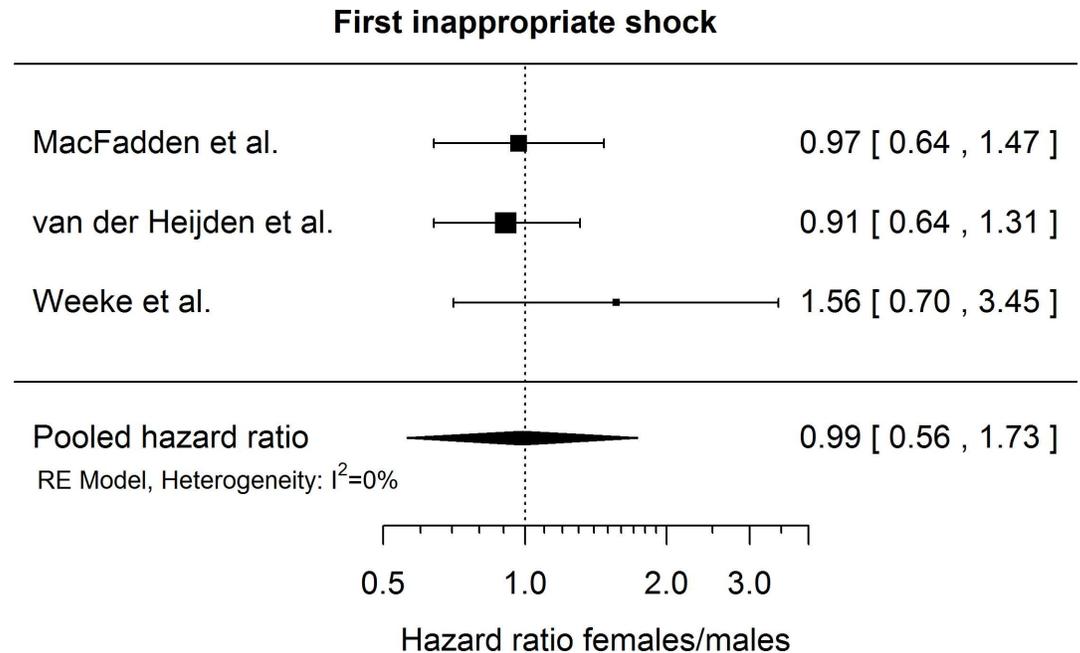

**Fig 4. Extracted hazard ratios for female gender regarding risk of inappropriate shocks with 95% confidence intervals as reported in the respective publications.** The pooled estimate is reported with a Knapp-Hartung adjusted 95% confidence interval. The dotted vertical line denotes a hazard ratio of 1, which corresponds to no difference in the risk between males and females.

doi:10.1371/journal.pone.0162756.g004

The current analysis should also be interpreted in the context of its limitations. First, as we intentionally focused on recent publications, the included studies were all observational, and the causality of the observed associations is uncertain. Second, this is a study level meta-analysis, as individual patient data were not available for most studies. Finally, it is important to note that our results do not apply to patients qualifying for secondary prophylactic ICD treatment.

In conclusion, in contemporary cohorts of patients receiving a primary prophylactic ICD, women have lower risks of death and appropriate shocks than men, but a similar risk of inappropriate shocks. These findings were observed in studies with a relatively low 2-year mortality. Our data suggest that further studies are warranted to validate the described gender-related differences and to defined improved risk stratification tools for primary prophylactic ICD implantation in patients with heart failure.

## Supporting Information

**S1 PRISMA Checklist. PRISMA checklist of items to include when reporting a systematic review or meta-analysis.**
(DOC)

**S1 Table. Sensitivity analyses for all three endpoints removing one study at a time**
(DOCX)

**S2 Table. Estimates for the cumulative incidence of death within the first 2 years after ICD implantation**
(DOCX)

## Acknowledgments

The authors gratefully acknowledge support by Dr Klaus Jung (Göttingen) in the reanalysis of the study by Seegers et al.

## Author Contributions

**Conceptualization:** MZ TF DC.

**Data curation:** BA CR.

**Formal analysis:** BA TF.

**Funding acquisition:** MZ TF DC CS.

**Investigation:** DC BA LB PM MZ SW.

**Methodology:** BA TF.

**Project administration:** BA.

**Resources:** MZ TF DC CS.

**Software:** BA TF.

**Supervision:** MZ TF DC.

**Validation:** CR LB PM.

**Visualization:** BA.

**Writing – original draft:** DC BA.

**Writing – review & editing:** MZ TF CR LB PM CS SW.

## References

1. Bardy GH, Lee KL, Mark DB, Poole JE, Packer DL, Boineau R, et al. Amiodarone or an implantable cardioverter-defibrillator for congestive heart failure. The New England journal of medicine. 2005; 352 (3):225–37. doi: 10.1056/NEJMoa043399 PMID: 15659722.

2. Moss AJ, Zareba W, Hall WJ, Klein H, Wilber DJ, Cannom DS, et al. Prophylactic implantation of a defibrillator in patients with myocardial infarction and reduced ejection fraction. The New England journal of medicine. 2002; 346(12):877–83. doi: 10.1056/NEJMoa013474 PMID: 11907286.

3. Kadish A, Dyer A, Daubert JP, Quigg R, Estes NA, Anderson KP, et al. Prophylactic defibrillator implantation in patients with nonischemic dilated cardiomyopathy. The New England journal of medicine. 2004; 350(21):2151–8. doi: 10.1056/NEJMoa033088 PMID: 15152060.

4. Al-Khatib SM, Anstrom KJ, Eisenstein EL, Peterson ED, Jollis JG, Mark DB, et al. Clinical and economic implications of the Multicenter Automatic Defibrillator Implantation Trial-II. Annals of internal medicine. 2005; 142(8):593–600. PMID: 15838065.

5. Mark DB, Nelson CL, Anstrom KJ, Al-Khatib SM, Tsiatis AA, Cowper PA, et al. Cost-effectiveness of defibrillator therapy or amiodarone in chronic stable heart failure: results from the Sudden Cardiac Death in Heart Failure Trial (SCD-HeFT). Circulation. 2006; 114(2):135–42. doi: 10.1161/CIRCULATIONAHA.105.581884 PMID: 16818817.

6. Loh JC, Creaser J, Rourke DA, Livingston N, Harrison TK, Vandenbogaart E, et al. Temporal trends in treatment and outcomes for advanced heart failure with reduced ejection fraction from 1993–2010: findings from a university referral center. Circulation Heart failure. 2013; 6(3):411–9. doi: 10.1161/CIRCHEARTFAILURE.112.000178 PMID: 23479563; PubMed Central PMCID: PMC3674961.

7. Gatzoulis KA, Tsiachris D, Dilaveris P, Archontakis S, Arsenos P, Vouliotis A, et al. Implantable cardioverter defibrillator therapy activation for high risk patients with relatively well preserved left ventricular ejection fraction. Does it really work? International journal of cardiology. 2013; 167(4):1360–5. doi: 10.1016/j.ijcard.2012.04.005 PMID: 22534047.

8. van der Heijden AC, Thijssen J, Borleffs CJ, van Rees JB, Hoke U, van der Velde ET, et al. Gender-specific differences in clinical outcome of primary prevention implantable cardioverter defibrillator recipients. Heart. 2013; 99(17):1244–9. doi: 10.1136/heartjnl-2013-304013 PMID: 23723448.

9. Weeke P, Johansen JB, Jorgensen OD, Nielsen JC, Moller M, Videbaek R, et al. Mortality and appropriate and inappropriate therapy in patients with ischaemic heart disease and implanted cardioverter-defibrillators for primary prevention: data from the Danish ICD Register. Europace: European pacing, arrhythmias, and cardiac electrophysiology: journal of the working groups on cardiac pacing, arrhythmias, and cardiac cellular electrophysiology of the European Society of Cardiology. 2013; 15(8):1150–7. doi: 10.1093/europace/eut017 PMID: 23407630.

10. Ghanbari H, Dalloul G, Hasan R, Daccarett M, Saba S, David S, et al. Effectiveness of implantable cardioverter-defibrillators for the primary prevention of sudden cardiac death in women with advanced heart failure: a meta-analysis of randomized controlled trials. Arch Intern Med. 2009; 169(16):1500–6. doi: 10.1001/archinternmed.2009.255 PMID: 19752408.

11. Santangeli P, Pelargonio G, Dello Russo A, Casella M, Bisceglia C, Bartoletti S, et al. Gender differences in clinical outcome and primary prevention defibrillator benefit in patients with severe left ventricular dysfunction: a systematic review and meta-analysis. Heart rhythm: the official journal of the Heart Rhythm Society. 2010; 7(7):876–82. doi: 10.1016/j.hrthm.2010.03.042 PMID: 20380893.

12. Wijers SC, van der Kolk BY, Tuinenburg AE, Doevendans PA, Vos MA, Meine M. Implementation of guidelines for implantable cardioverter-defibrillator therapy in clinical practice: Which patients do benefit? Netherlands heart journal: monthly journal of the Netherlands Society of Cardiology and the Netherlands Heart Foundation. 2013; 21(6):274–83. doi: 10.1007/s12471-013-0407-x PMID: 23572330; PubMed Central PMCID: PMC3661880.

13. Seegers J, Conen D, Jung K, Bergau L, Dorenkamp M, Luthje L, et al. Sex difference in appropriate shocks but not mortality during long-term follow-up in patients with implantable cardioverter-defibrillators. Europace: European pacing, arrhythmias, and cardiac electrophysiology: journal of the working groups on cardiac pacing, arrhythmias, and cardiac cellular electrophysiology of the European Society of Cardiology. 2015. doi: 10.1093/europace/euv361 PMID: 26622054.

14. Sanderson S, Tatt ID, Higgins JP. Tools for assessing quality and susceptibility to bias in observational studies in epidemiology: a systematic review and annotated bibliography. Int J Epidemiol. 2007; 36(3):666–76. doi: 10.1093/ije/dym018 PMID: 17470488.

15. Tierney JF, Stewart LA, Ghersi D, Burdett S, Sydes MR. Practical methods for incorporating summary time-to-event data into meta-analysis. Trials. 2007; 8:16. doi: 10.1186/1745-6215-8-16 PMID: 17555582; PubMed Central PMCID: PMC1920534.

16. Knapp G, Hartung J. Improved tests for a random effects meta-regression with a single covariate. Statistics in medicine. 2003; 22(17):2693–710. doi: 10.1002/sim.1482 PMID: 12939780.

17. Viechtbauer W. Conducting Meta-Analyses in R with the metafor Package. J Stat Softw. 2010; 36(3):1–48. PMID: WOS:000281593200001.

18. Schmidt M, Pedersen SB, Farkas DK, Hjortshoj SP, Botker HE, Nielsen JC, et al. Thirteen-year nationwide trends in use of implantable cardioverter-defibrillators and subsequent long-term survival. Heart rhythm: the official journal of the Heart Rhythm Society. 2015; 12(9):2018–27. doi: 10.1016/j.hrthm.2015.04.040 PMID: 25937527.

19. Bilchick KC, Stukenborg GJ, Kamath S, Cheng A. Prediction of mortality in clinical practice for medicare patients undergoing defibrillator implantation for primary prevention of sudden cardiac death. Journal of the American College of Cardiology. 2012; 60(17):1647–55. doi: 10.1016/j.jacc.2012.07.028 PMID: 23021331; PubMed Central PMCID: PMC3677038.

20. Campbell NG, Cantor EJ, Sawhney V, Duncan ER, DeMartini C, Baker V, et al. Predictors of new onset atrial fibrillation in patients with heart failure. International journal of cardiology. 2014; 175(2):328–32. doi: 10.1016/j.ijcard.2014.05.023 PMID: 24985070.

21. Demirel F, Adiyaman A, Timmer JR, Dambrink JH, Kok M, Boeve WJ, et al. Myocardial scar characteristics based on cardiac magnetic resonance imaging is associated with ventricular tachyarrhythmia in patients with ischemic cardiomyopathy. International journal of cardiology. 2014; 177(2):392–9. doi: 10.1016/j.ijcard.2014.08.132 PMID: 25440471.

22. Hage FG, Aljaroudi W, Aggarwal H, Bhatia V, Miller J, Doppalapudi H, et al. Outcomes of patients with chronic kidney disease and implantable cardiac defibrillator: primary versus secondary prevention. International journal of cardiology. 2013; 165(1):113–6. doi: 10.1016/j.ijcard.2011.07.087 PMID: 21862150.

23. Gigli L, Barabino D, Sartori P, Rossi P, Reggiardo G, Chiarella F, et al. The implantable cardioverter defibrillator in primary prevention: a revision of monocentric study group. Journal of cardiovascular medicine. 2014; 15(8):653–8. doi: 10.2459/JCM.0000000000000112 PMID: 24983347.

24. Kraaier K, Scholten MF, Tijssen JG, Theuns DA, Jordaens LJ, Wilde AA, et al. Early mortality in prophylactic implantable cardioverter-defibrillator recipients: development and validation of a clinical risk score. Europace: European pacing, arrhythmias, and cardiac electrophysiology: journal of the working groups on cardiac pacing, arrhythmias, and cardiac cellular electrophysiology of the European Society of Cardiology. 2014; 16(1):40–6. doi: 10.1093/europace/eut223 PMID: 23918791.

25. Levine YC, Rosenberg MA, Mittleman M, Samuel M, Methachittiphan N, Link M, et al. B-type natriuretic peptide is a major predictor of ventricular tachyarrhythmias. Heart rhythm: the official journal of the Heart Rhythm Society. 2014; 11(7):1109–16. doi: 10.1016/j.hrthm.2014.04.024 PMID: 24837348.

26. MacFadden DR, Crystal E, Krahn AD, Mangat I, Healey JS, Dorian P, et al. Sex differences in implantable cardioverter-defibrillator outcomes: findings from a prospective defibrillator database. Annals of internal medicine. 2012; 156(3):195–203. doi: 10.7326/0003-4819-156-3-201202070-00007 PMID: 22312139.

27. Smith T, Theuns DA, Caliskan K, Jordaens L. Long-term follow-up of prophylactic implantable cardioverter-defibrillator-only therapy: comparison of ischemic and nonischemic heart disease. Clinical cardiology. 2011; 34(12):761–7. doi: 10.1002/clc.20970 PMID: 22038531.

28. Yung D, Birnie D, Dorian P, Healey JS, Simpson CS, Crystal E, et al. Survival after implantable cardioverter-defibrillator implantation in the elderly. Circulation. 2013; 127(24):2383–92. doi: 10.1161/CIRCULATIONAHA.113.001442 PMID: 23775193.

29. Providencia R, Marijon E, Lambiase PD, Bouzeman A, Defaye P, Klug D, et al. Primary Prevention Implantable Cardioverter Defibrillator (ICD) Therapy in Women-Data From a Multicenter French Registry. J Am Heart Assoc. 2016; 5(2). doi: 10.1161/JAHA.115.002756 PMID: 26873687; PubMed Central PMCID: PMC4802475.

30. Stabile G, D'Agostino C, Gallo P, Marrazzo N, Iuliano A, De Simone A, et al. Appropriate therapies predict long-term mortality in primary and secondary prevention of sudden cardiac death. Journal of cardiovascular medicine. 2013; 14(2):110–3. doi: 10.2459/JCM.0b013e3283511f5b PMID: 22367567.

31. Amit G, Suleiman M, Konstantino Y, Luria D, Kazatsker M, Chetboun I, et al. Sex differences in implantable cardioverter-defibrillator implantation indications and outcomes: lessons from the Nationwide Israeli-ICD Registry. Europace: European pacing, arrhythmias, and cardiac electrophysiology: journal of the working groups on cardiac pacing, arrhythmias, and cardiac cellular electrophysiology of the European Society of Cardiology. 2014; 16(8):1175–80. doi: 10.1093/europace/euu015 PMID: 24554524.

32. Masoudi FA, Go AS, Magid DJ, Cassidy-Bushrow AE, Gurwitz JH, Liu TI, et al. Age and sex differences in long-term outcomes following implantable cardioverter-defibrillator placement in contemporary clinical practice: findings from the Cardiovascular Research Network. J Am Heart Assoc. 2015; 4(6): e002005. doi: 10.1161/JAHA.115.002005 PMID: 26037083; PubMed Central PMCID: PMC4599538.

33. Rodriguez-Manero M, Barrio-Lopez MT, Assi EA, Exposito-Garcia V, Bertomeu-Gonzalez V, Sanchez-Gomez JM, et al. Primary Prevention of Sudden Death in Patients With Valvular Cardiomyopathy. Rev Esp Cardiol (Engl Ed). 2016; 69(3):272–8. doi: 10.1016/j.rec.2015.05.016 PMID: 26481284.

34. MERIT-HF Study Group. Effect of metoprolol CR/XL in chronic heart failure: Metoprolol CR/XL Randomised Intervention Trial in Congestive Heart Failure (MERIT-HF). Lancet. 1999; 353(9169):2001–7. PMID: 10376614.

35. Pitt B, Zannad F, Remme WJ, Cody R, Castaigne A, Perez A, et al. The effect of spironolactone on morbidity and mortality in patients with severe heart failure. Randomized Aldactone Evaluation Study Investigators. The New England journal of medicine. 1999; 341(10):709–17. doi: 10.1056/NEJM199909023411001 PMID: 10471456.

36. Koller MT, Schaer B, Wolbers M, Sticherling C, Bucher HC, Osswald S. Death without prior appropriate implantable cardioverter-defibrillator therapy: a competing risk study. Circulation. 2008; 117(15):1918–26. doi: 10.1161/CIRCULATIONAHA.107.742155 PMID: 18391108.